\documentclass[prb,twocolumn,showpacs]{revtex4}

\usepackage{graphicx}

\begin{document}

\author{O. Leenaerts}
\email{ortwin.leenaerts@ua.ac.be} \affiliation{Universiteit Antwerpen, Departement Fysica,
Groenenborgerlaan 171, B-2020 Antwerpen, Belgium}
\author{B. Partoens}
\email{bart.partoens@ua.ac.be} \affiliation{Universiteit Antwerpen, Departement Fysica,
Groenenborgerlaan 171, B-2020 Antwerpen, Belgium}
\author{F. M. Peeters}
\email{francois.peeters@ua.ac.be} \affiliation{Universiteit Antwerpen, Departement Fysica,
Groenenborgerlaan 171, B-2020 Antwerpen, Belgium}
\date{\today}
\title{Paramagnetic adsorbates on graphene: a charge transfer analysis}

\begin{abstract}
We introduce a modified version of the Hirshfeld charge analysis method and demonstrate its accurateness 
by calculating the charge transfer between the paramagnetic molecule NO$_2$ and
graphene. The charge transfer between paramagnetic molecules and a graphene layer as calculated with 
ab initio methods can crucially depend on the size of the supercell
used in the calculation. This has important consequences for adsorption studies involving
paramagnetic molecules such as NO$_2$ physisorbed on graphene or on carbon nanotubes.
\end{abstract}

 \maketitle

The use of carbon nanotubes\cite{kong,li,bradley,qi,staii,tang} and more recently
graphene\cite{schedin} as very sensitive gas sensors has stimulated a lot of theoretical work on
this subject.\cite{peng,santucci,peng2,zhao,hwang,wehling,leenaerts,zanella} Through ab initio
calculations, one investigated the adsorption properties of these carbon materials and looked for
the mechanisms behind their good sensing capabilities. It appears that the key issue is
the charge transfer from the gas molecules to the carbon surface. Such ab initio calculations were able
to provide good qualitative agreement with experiment, e.g. whether the gas molecules act as electron
donors or acceptors, but a large variation in the size of the doping was found between different
theoretical calculations. For example, the calculated charge transfer between a (10,0) single-walled carbon nanotube (SWNT) and a NO$_2$ molecule varies from\cite{santucci} $-0.015e$ to\cite{peng2} $-0.10e$, an order of magnitude
difference! The size of the doping is, however, a crucial factor determining the sensitivity of
the gas sensor.  In this letter we demonstrate that important reasons for the discrepancies
are: i) the different sizes of the supercells used in these ab initio calculations, and ii) the theoretical
charge analysis method that has been used. We will show that the former is a very decisive parameter
when the adsorbing molecules are paramagnetic. First we examine critically different theoretical charge
analysis methods. Then we will perform ab initio calculations with different supercells to investigate
the charge transfer dependence on the size of the supercell.

All our ab initio calculations were done within the density funcional theory (DFT) formalism using the ABINIT\cite{abinit}
software package within the local spin density approximation (LSD) and with Troullier-Martins pseudopotentials.\cite{troullier} We used a plane wave basis set
with an energy cutoff of 816 eV, which was tested to give converged results for all the properties
studied in this letter. Different graphene supercells ranging from $2\times 2$ to $6\times 6$ were
implemented. For the sampling of the Brillouin zone (BZ) we used  Monckhorts-Pack (MP) grids for
the different supercells equivalent to a range from $12\times 12\times 1$ to $48\times 48\times 1$
points for a single unit cell.

To calculate the charge transfer from a molecule to a surface, one needs a physically meaningful
and transparent approach to divide the electron density between them. A variety of methods has
been developed for this purpose, such as the Mulliken's\cite{mulliken} or Bader's\cite{bader}
atoms in molecules approaches.  We introduce a method based on the Hirshfeld\cite{hirshfeld}
method and in particular the recently extended version of this approach, the
Hirshfeld-I\cite{alsenoy} method. The simplicity and low cost of computation time makes the
Hirshfeld method a very powerfull approach in DFT calculations, and we will demonstrate that its
extended version is more accurate than the Bader method and other Hirshfeld-based methods. In the
Hirshfeld charge analysis, the total electron density $\rho$ is divided between the different
atoms of a system, according to the density of the neutral atoms, $\rho_{A}^{0}$, in free space
which build up this system: $Q_A =\int{ (\rho_A^0({\mathbf r})/\sum_{A'} \rho_{A'}^0({\mathbf r})
}) \rho({\mathbf r})d{\mathbf r}$, with $Q_{A}$ the charge on atom $A$ (in units of $e$). The
Hirshfeld-I method does not use the density of neutral atoms in free space, but uses the density
of charged atoms instead. These charges on the different atoms are determined through an iterative
process. One starts with the simple Hirshfeld method and determines the charge per atom. Then this
charge is placed on the different atoms in free space and their densities are used to divide the
density again among the atoms as with the simple Hirshfeld method. This procedure is repeated
until the charges on the atoms are converged. In our approach we allow our system to relax
completely and calculate the total density. Then we calculate the density of the molecule and the
graphene layer separately in the same configuration as in the total relaxed system. These
densities can now be used in the Hirshfeld-I method instead of the separate atomic densities. Such
an approach gives charge transfers that are more physically meaningful as we will demonstrate in
the following.

To test this charge analysis procedure we investigate the adsorption of the paramagnetic NO$_2$
molecule (with magnetic moment $M=1\mu_b$) on a $4\times4$ graphene supercell. We calculate the charge transfer of the
NO$_2$-graphene system using different theoretical approaches: we first use the simple Hirshfeld
method, then we change the atomic densities into a molecular and a graphene density and use the
same procedure; finally we put charges on NO$_2$ and graphene in order to use them in the
Hirshfeld-I charge analysis. The Hirshfeld methods that use molecular densities
instead of atomic densities will be referred to as the modified Hirshfeld and modified Hirshfeld-I method. For
comparison, we also calculated the charge transfer based on the Bader charge analysis. The charge
transfer results are given in table~\ref{tab-1}.

\begin{table}[h]
\caption{The charge transfer from graphene to NO$_2$ calculated with different methods.\label{tab-1}}
\begin{tabular}{cc}
\hline\hline
method & { charge transfer ($e$)}  \\
\hline

       Hirshfeld            &  -0.099     \\

       modified Hirshfeld   &  -0.161     \\

       modified Hirshfeld-I &  -0.181     \\

       Bader                &  -0.212     \\

       magnetic moment      &  -0.182     \\

\hline\hline
\end{tabular}
\end{table}

The advantage of using a paramagnetic molecule to demonstrate the accurateness of this charge
transfer analysis method is that we can have a good estimate of the charge on the molecule through
some physically relevant properties of the system, like the magnetic moment: extra charge in the
partially occupied molecular orbital (POMO) of NO$_2$ will lower the magnetic moment of the NO$_2$
molecule. Extra charge on the graphene layer, on the other hand, will not effect its magnetic moment 
because pure graphene is diamagnetic\cite{mcclure}. This implies that we can simply use the lowering of 
the magnetic moment of the total system (i.e. the difference from 1$\mu_b$) to estimate the charge 
transfer between the NO$_2$ molecule and graphene. This is very accurate if there is
not too much hybridization between the molecular orbitals of NO$_2$ and the graphene orbitals.
This is indeed the case for NO$_2$ since it physisorbs on graphene.\cite{wehling,leenaerts} The
value of the charge transfer extracted from the lowering of the magnetic moment is $-0.182e$ (see
table~\ref{tab-1}), which is very close to the charge transfer obtained from the modified
Hirshfeld-I method. The use of molecular densities instead of atomic ones leads already to a
significant improvement of the simple Hirshfeld method and the iterative procedure makes the
calculated charge transfer almost equal to the one extracted from the change in magnetic moment of
the total system. The small difference of the order of $0.001$e is caused by a small charge
transfer due to orbital hybridization. This causes no change in the magnetic moment, but it is
noticeable in the modified Hirshfeld-I charge analysis. Note also that the charge transfer from
the Bader analysis does not correspond with the charge transfer from the lowering of the magnetic
moment (see table~\ref{tab-1}). Thus the modified Hirshfeld-I method gives a good approximation to
the charge transfer in the adsorption process of paramagnetic molecules on graphene and, if the
molecules are not paramagnetic, it is probably also more accurate to use this modified Hirshfeld-I
charge analysis instead of e.g the Bader charge analysis for a trustworthy determination of the
charge transfer.

\begin{figure}[h]
  \centering
\includegraphics[width=3.4in]{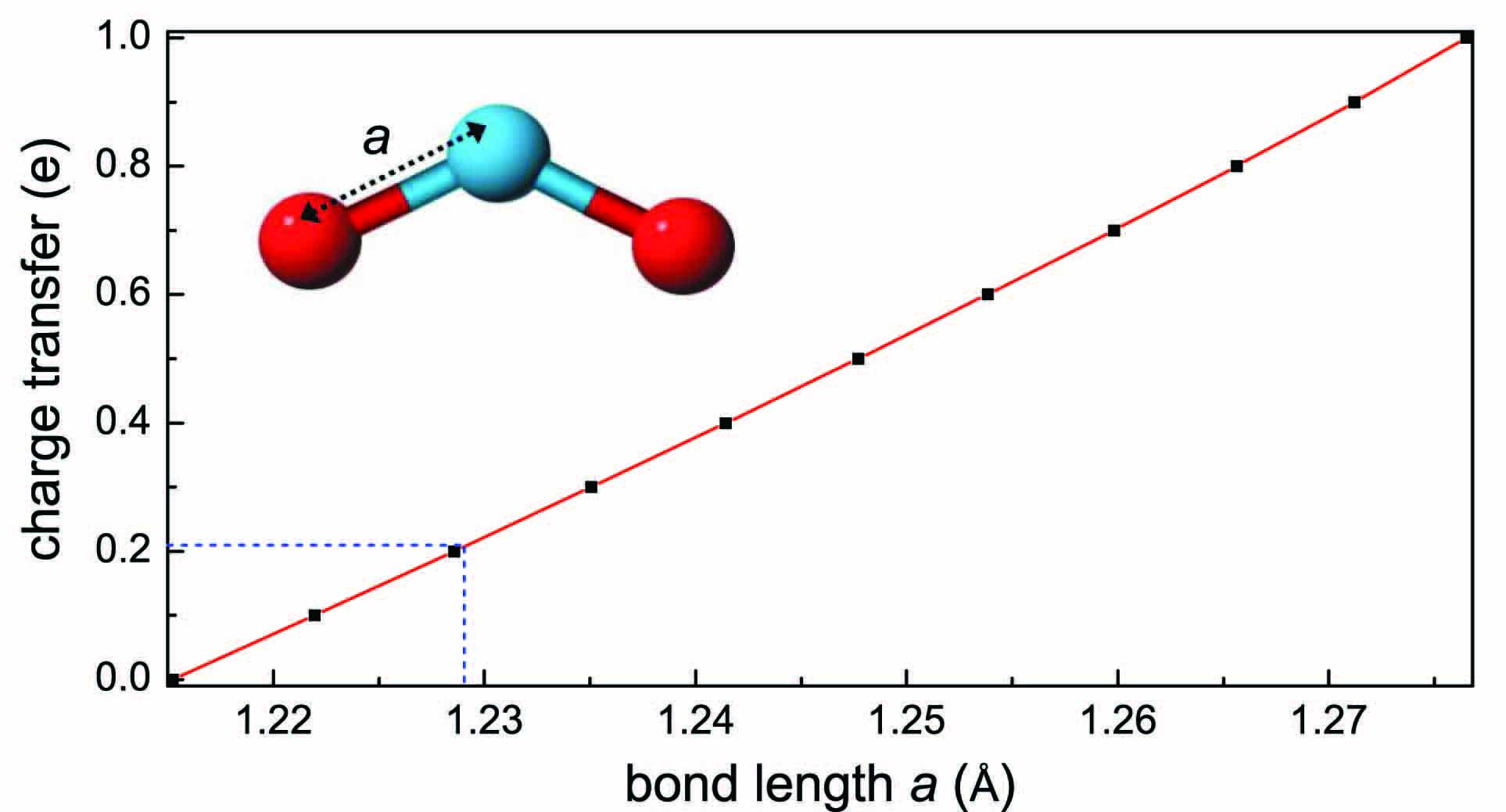}
\caption{\label{fig1}(Color online) The dependence of the bond length of a NO$_2$ molecule on the
extra charge put on the molecule.}
\end{figure}

Another element of support for the modified Hirshfeld-I method is its geometrical interpretation.
When we put an extra charge on a free standing paramagnetic molecule, it will be placed at the
POMO. In the case of a NO$_2$ molecule, this is an anti-bonding orbital which means that the bond
length between the N and O atoms becomes larger when the POMO becomes filled. This fact can be
used to estimate the charge transfer of a physisorbed NO$_2$ molecule to graphene
by comparing the bond length of the relaxed physisorbed molecule with a charged free standing
NO$_2$ molecule (see Fig.~\ref{fig1}). From Fig.~\ref{fig1} we extract a charge transfer of
$-0.206e$, which is close to the charge transfer obtained by the
modified Hirshfeld-I method (and not close to the simple and modified Hirshfeld methods).\\

\begin{figure}[h]
  \centering
\includegraphics[width= 3.4 in]{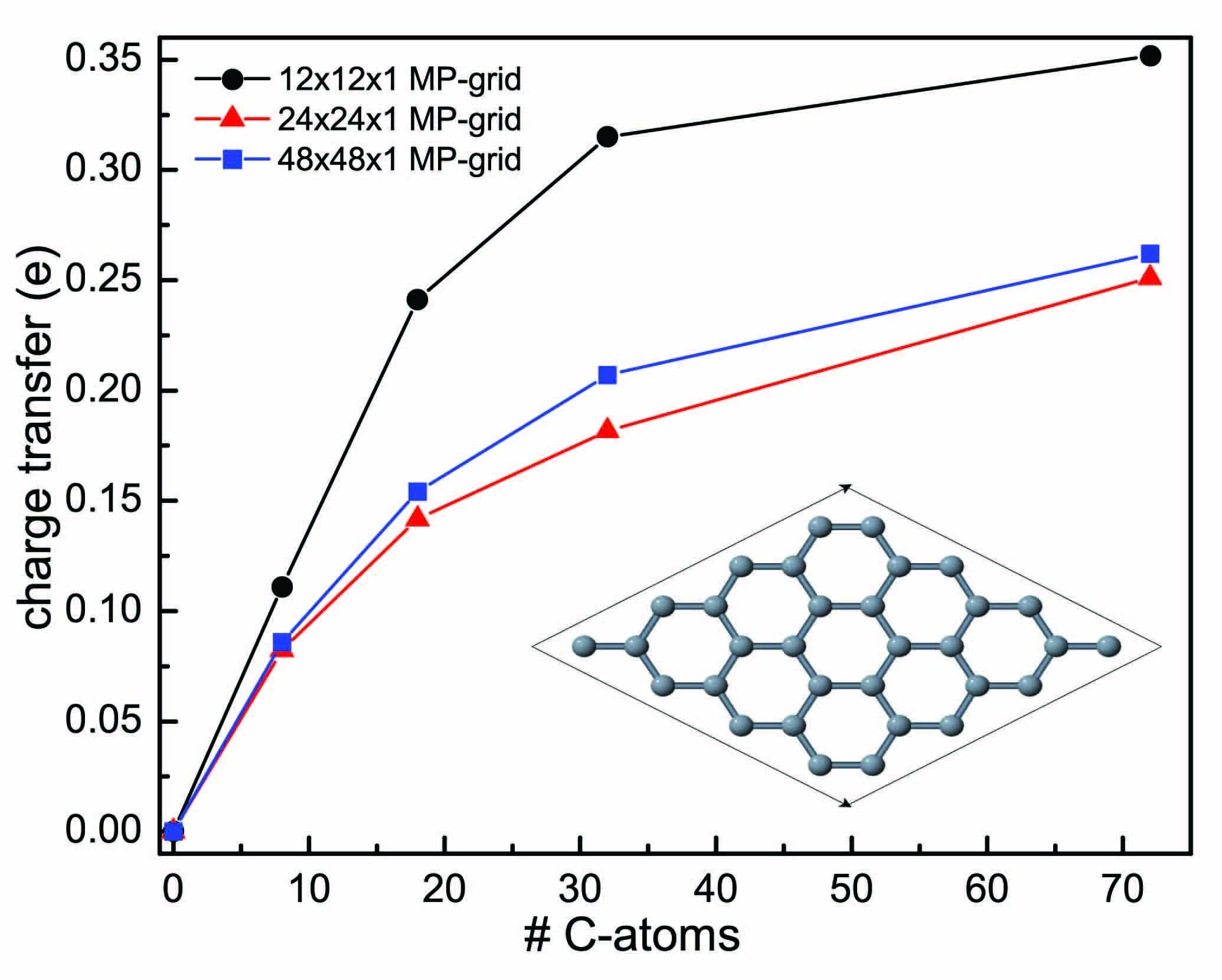}
\caption{\label{fig2}(Color online) Supercell dependence of the charge transfer between a NO$_2$ 
molecule and graphene for different MP-grids. The inset shows a $4\times4$ graphene supercell.}
\end{figure}

Next we will discuss how the charge transfer for the case of paramagnetic molecules depends on the
size of the supercell used in the ab initio calculations. The simulation of adsorption processes
at surfaces is necessarily restricted to finite supercells. In most cases this gives a reasonably
good approximation for e.g. charge transfer analyses in physisorption processes, but in some
cases, when e.g. the adsorbates are paramagnetic, a different kind of charge transfer mechanism
dominates the charge transfer\cite{wehling,leenaerts}, leading to very different results. We will
use the NO$_2$ molecule again to show how sensitive the charge transfer can be with respect to the
size of the used supercell. We make use of 4 different supercells, $2\times2, 3\times3, 4\times4,$
and $6\times6$ and several MP-grids for sampling the BZ. Fig.~\ref{fig2} shows that the charge
transfers are converged reasonably well for a $24\times24\times1$ MP-grid.

\begin{figure}[h]
  \centering
\includegraphics[width= 3.4 in]{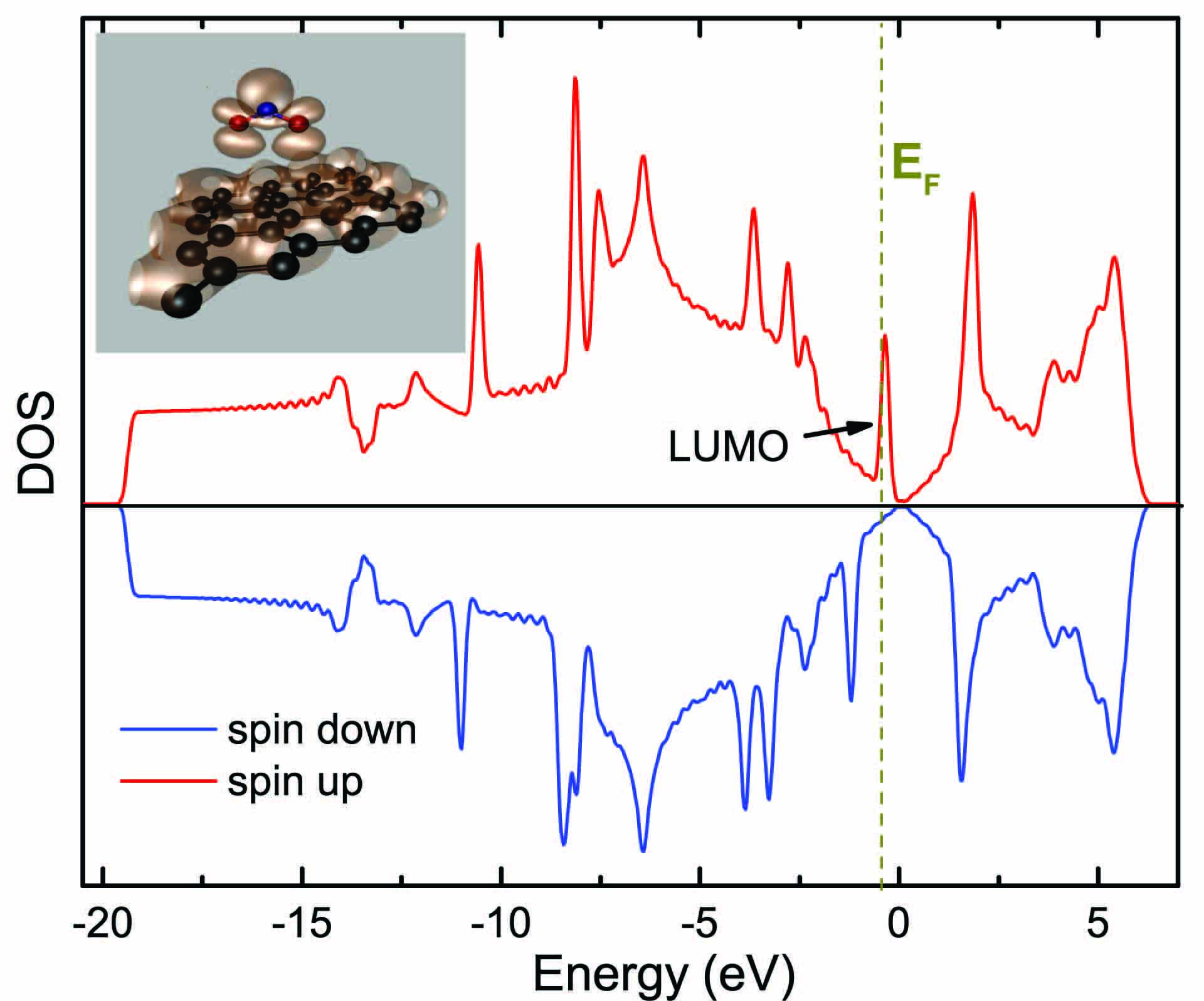}
\caption{\label{fig3}(Color online) Spin-polarized DOS for NO$_2$ adsorbed on graphene calculated
with a $60\times60\times1$ MP-grid for a $4\times4$ supercell. The LUMO of the NO$_2$
molecule and the Fermi-energy are indicated. The inset shows the orbitals of the NO$_2$ molecule
(the LUMO) and graphene between which the electron transfer takes place.}
\end{figure}

\begin{figure}[h!]
  \centering
\includegraphics[width= 3.4 in]{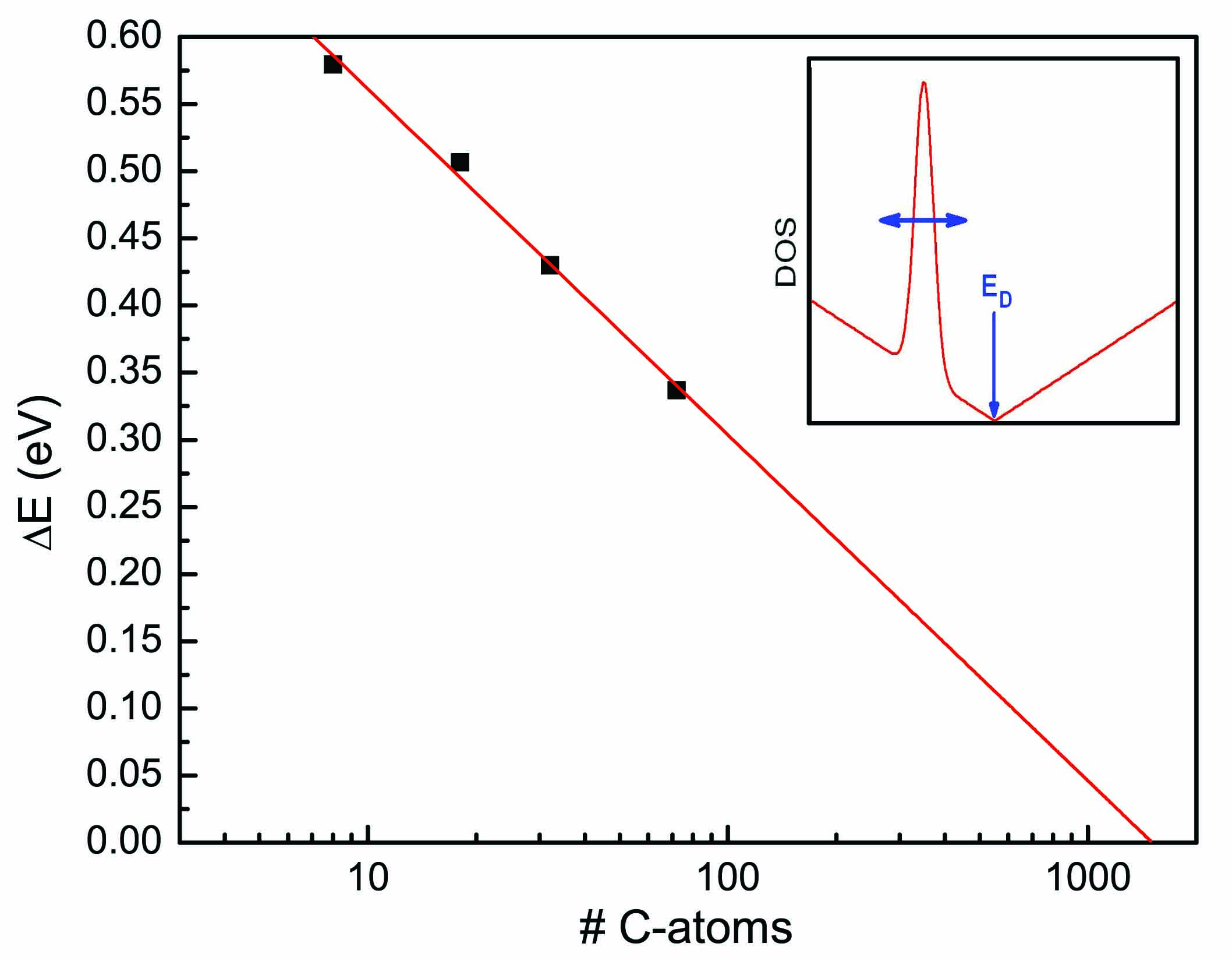}
\caption{\label{fig4}(Color online) Distance, $\Delta$E, of the NO$_2$ LUMO to the Dirac-point with
respect to the number of carbon atoms used in the supercell. The inset shows a close-up of
the total DOS of graphene with an adsorbed NO$_2$ molecule around the Dirac-point.}
\end{figure}

There is a pronounced dependence of the charge transfer on the number of atoms in the simulated
graphene layer\cite{number} which can be explained by looking at the density of states (DOS) of
the total system\cite{wehling,leenaerts} (see Fig.~\ref{fig3}). The lowest unoccupied orbital
(LUMO) of NO$_2$ is below the Dirac-point\cite{dirac} of graphene, which causes a charge transfer
to the molecule. This charge transfer depends clearly on the number of electronic states between
the Dirac-point and the LUMO of the NO$_2$ molecule, which depends linearly on the number of
carbon atoms in the supercell. Therefore, we may expect that if one takes a graphene supercell that is large enough, one
would eventually get a charge transfer of one electron.\cite{wehling}  However, it is clear from
Fig.~\ref{fig2} that there is no simple linear dependence. The position of the LUMO in the DOS
depends on the filling of this orbital: the more electrons in the orbital the more difficult it
gets to put another one in it and this translates in a shift of the LUMO towards the Dirac-point.
In Fig.~\ref{fig4} we show this shift as a function of the number of carbon atoms in the supercell. It is not obvious that
the orbital becomes totally filled (a transfer of one electron per paramagnetic molecule) before it
coincides with the Dirac point, where the shift will stop and the charge transfer
will be converged at a value of less than one electron per molecule.

Thus one can only obtain a quantitative meaningful value of the charge transfer between a
paramagnetic molecule and a graphene layer in two cases: i) when the supercell used is large
enough to get a charge transfer of one electron per molecule, or ii) the LUMO of the molecule
coincides with the Dirac-point and the charge transfer has converged. In the case of the NO$_2$
molecule, one needs a supercell that is much larger than the $6\times6$ supercell we
were able to use due to computational limitations. However, from Fig.~\ref{fig4} we find that if the LUMO 
stays partially full, it takes more
than thousand carbon atoms to let it coincide with the Dirac-point. On the other hand, from Fig.~\ref{fig2} one
may deduce that it is not unlikely that a charge transfer of $1e$ is reached before the supercell
contains 1000 carbon atoms. From this we can conclude that probably the LUMO of NO$_2$ will be completely filled
before it reaches the Dirac-point. This is compatible with the experimental
observation\cite{schedin} that the charge transfer equals one electron charge per NO$_2$ molecule.\\

To conclude, we examined different methods to calculate the charge transfer between a paramagnetic
molecule and a graphene layer and found that a modified Hirshfeld-I method gives the most accurate
and physically meaningful results. We used this method to investigate the dependence of the charge transfer
on the size of the supercell in ab initio calculations and found that charge transfers involving
paramagnetic molecules are, in a nontrivial way, very sensitive to the supercell size. We applied our results to the 
adsorption of NO$_2$ on graphene and showed that our ab initio calculations are compatible with
the experimentally found charge transfer of $1e$.\cite{schedin} In contrast to the claim in Ref.~\onlinecite{wehling}, 
we found that a charge transfer of $1e$ is not always realized for adsorption of a paramagnetic molecules on graphene. 
The present results are also valid for single wall carbon nanotubes because their adsorption properties are
similar to those of graphene.\cite{leenaerts}

\begin{acknowledgments}
This work was supported by the Flemish Science Foundation (FWO-Vl), the NOI-BOF of the University
of Antwerp and the Belgian Science Policy (IAP). Discussions with C. Van Alsenoy are gratefully
acknowledged.
\end{acknowledgments}


\begin{thebibliography}{99}
\bibitem{number}The number of atoms in the simulated graphene layer is equal to twice the number of unit cells in the supercell.
\bibitem{dirac} The Dirac-point is the Fermi-level of pure graphene.
\end{thebibliography}
\end{document}